# Exploring the substrate-driven morphological changes in Nd$_{0.6}$Sr$_{0.4}$MnO$_3$ thin films

R S Mrinaleni [1,2], E P Amaladass[1,2*], S Amirthapandian [1,2], A. T. Sathyanarayana [1,2], Jegadeesan P [1,2], Ganesan K [1,2], R M Sarguna [1,2], P. N. Rao [3], Pooja Gupta [3,4], T Geetha Kumary[1,2], and S. K. Rai [3,4], Awadhesh Mani[1,2]

[1]Material Science Group, Indira Gandhi Centre for Atomic Research, Kalpakkam, 603102, India
[2]Homi Bhabha National Institute, Indira Gandhi Centre for Atomic Research, Kalpakkam 603102, India
[3]Synchrotrons Utilisation Section, Raja Ramanna Centre for Advanced Technology, PO RRCAT, Indore, Madhya Pradesh 452013, India
[4]Homi Bhabha National Institute, Training School Complex, Anushaktinagar, Mumbai, Maharashtra 400094, India

*Corresponding author: edward@igcar.gov.in

## ABSTRACT

Manganite thin films are promising candidates for studying the strongly correlated electron systems. Understanding the growth-and morphology-driven changes in the physical properties of manganite thin films is vital for their applications in oxitronics. This work reports the morphological, structural, and electrical transport properties of nanostructured Nd$_{0.6}$Sr$_{0.4}$MnO$_3$ (NSMO) thin films fabricated using the pulsed laser deposition technique. Scanning electron microscopy (SEM) imaging of the thin films revealed two prominent surface morphologies: a granular and a unique crossed-nano-rod-type morphology. From X-ray diffraction (XRD) and atomic force microscopy (AFM) analysis, we found that the observed nanostructures resulted from altered growth modes occurring on the terraced substrate surface. Furthermore, investigations on the electrical-transport properties of thin films revealed that the films with crossed-nano-rod type morphology showed a sharp resistive transition near the metal-to-insulator transition (MIT). An enhanced temperature coefficient of resistance (TCR) of up to one order of magnitude was also observed compared to the films with granular morphology. Such enhancement in TCR % by tuning the morphology makes these thin films promising candidates for developing oxide-based temperature sensors and detectors.

**INTRODUCTION**

$Nd_{0.6}Sr_{0.4}MnO_3$ (NSMO) belongs to the class of magnetic oxides $RE_{1-x}A_xMnO_3$ (where RE= $La^{3+}$, $Nd^{3+}$, $Pr^{3+}$, $Sm^{3+}$, and A = $Ca^{2+}$, $Sr^{2+}$, $Ba^{2+}$, etc.) with perovskite ($ABO_3$) structure which exhibits a variety of magnetic phases by tuning the dopant concentration x (x = 0 to 0.9)[1–3]. Manganites are known for their exotic properties such as the Colossal magnetoresistive (CMR) phenomenon[4], Metal-insulator-transition (MIT) accompanied by a magnetic transition from paramagnetic (PM) to ferromagnetic (FM) state[5], half-metallicity[6], and tuneable in-plane and out of plane magnetic anisotropy[7]. These properties are exploited for potential spintronics applications such as spin injection devices[8], Magnetic tunnel junctions[9–11], and magnetic storage devices (MRAMs)[12]. In recent times, the perovskite-manganite systems are the ideal oxide candidates for developing superlattices, self-assembled nano-arrays[13], nano-ribbons[14], nano-wires, vertically aligned nanocomposite (VAN) thin films[15–19], etc. which offer enhanced Low-field magnetoresistance (LFMR), switchable magnetic anisotropy and for studying other interesting interface effects such as magnetic exchange bias[20]. Focus on growth dynamics is required to tune exclusive nano-architectures in the thin film as it offers additional handles to tailor its physical properties such as a high CMR %, high Curie & MIT temperature, high-temperature coefficient of resistance (TCR %), and enhanced magnetoresistive (MR) phenomenon. The manganite system is highly sensitive to external perturbations due to the strong connection between the spin-charge and lattice degrees of freedom[21,22]. This poses a major challenge in obtaining epitaxial/patterned thin films for useful applications.

The pulsed laser deposition (PLD) technique has been extensively used to fabricate oxide-based manganite thin films. This is because it offers good stoichiometric transfer of the target material onto the substrate in addition to deposition in an oxygen background. Various studies have been carried out to obtain epitaxial thin films by tuning the deposition parameters such as the oxygen partial pressure, substrate temperature, laser energy density, and repetition rate, affecting its growth and physical properties[23,24]. Additionally, the growth of the thin film is influenced by the substrate. The strain offered by the substrate affects the surface morphology and microstructure of the manganite thin film. Different methodologies such as i) varying the substrates for different lattice matching[25–27] (ii) choice of substrates with different crystallographic orientations with corresponding chemical terminations[14] iii) varying the thickness of the thin films[28], and iv) high-temperature annealing[17] are adopted to tune the strain and morphology of the thin films. Therefore, thin films with unique morphology and long-range ordered nanostructures can be obtained by fine-tuning the growth parameters. Compared to the previous works on VAN and other nanostructures of the popular manganite system La-Sr-Mn-O, we have observed a granular nanostructure and another distinct nanostructure with crossed-nano-rods in our thin films. We have synthesized NSMO thin films using the PLD technique on single-crystal SrTiO3 (100) oriented substrates (STO). The effects of PLD parameters and annealing conditions on the surface morphology were investigated. Using SEM, AFM, and XRD techniques, the growth mechanism leading to a specific type of nano-structuring in the NSMO thin films is studied. Additionally, the morphology-driven changes in the temperature dependence of resistivity are investigated, and we observed a signature trend in the MIT corresponding to the particular morphology.

**EXPERIMENTAL METHODS**

The NSMO thin films were fabricated using the PLD technique using a commercial NSMO pellet as the target. Before deposition, SrTiO$_3$ (STO) *(1 0 0)* single crystals substrate was cleaned by boiling in de-ionized(DI) water for 3 minutes, followed by ultra-sonication in DI water, acetone, and iso-propyl alcohol followed by rinsing in DI water. With the water leaching procedure, the SrO terminations present in the substrate surface can be effectively dissolved and removed with DI at elevated temperatures > 60 ºC followed by ultra-sonication. A KrF Excimer laser source ($\lambda$ = 248 nm) operated with a laser energy density of 1.75 J/cm$^2$ at 3Hz was used to ablate the target. The films were deposited in an oxygen partial of 0.36 mbar with substrate temperature fixed at 750 ºC. After deposition, the films were in situ annealed at 750 ºC for 2h, and the PLD chamber was maintained with O$_2$ background pressure of 0 to 1 bar. Further, the films were ex-situ annealed in a tube furnace at 950 ºC in an oxygen atmosphere with a flow rate of ~ 20 sccm for 2h.

The surface morphology of the thin films was examined using a Scanning electron microscope (SEM) from Carl Zeiss, crossbeam 340, and the images were collected in inlens-duo mode at 3-5 kV. Atomic force microscopy (AFM) was used for 2D and 3D visualization of the surface of substrates and the films. XRD studies have been carried out at Engineering Applications Beamline, BL-02, Indus-2 synchrotron source, India using beam energy of 15 keV for the structural characterization of the films[29]. The Grazing incidence (GI) and ω-2θ scans were performed, and data were collected using the Dectris detector (MYTHEN2 X 1K) in reflection geometry. In the GI-scan, the incident angle is kept fixed at ω = 0.5º, and the detector moves along the given 2θ range. The monochromatic high-resolution mode of the beamline was used, keeping the beam energy at 15 keV (λ = 0.826 Å). The peaks were indexed with reference to the ICDD data[30] (ICDD number - 01-085-6743)

## RESULTS AND DISCUSSION:

### 1. *Morphology studies of the nanostructured thin films:*

The NSMO thin films prepared under the above conditions possessed two prominent surface morphology – granular and rod-type. Two representative films with granular nanostructure and crossed-rod nanostructure were chosen to study the physical properties. These two systems will be referred to as NS-G and NS-R, where NS stands for NSMO thin film, and 'G'/'R' stands for the type of morphology. The thickness of NS-G and NS-R thin films is determined to be ~ 100 nm by cross-sectional SEM.

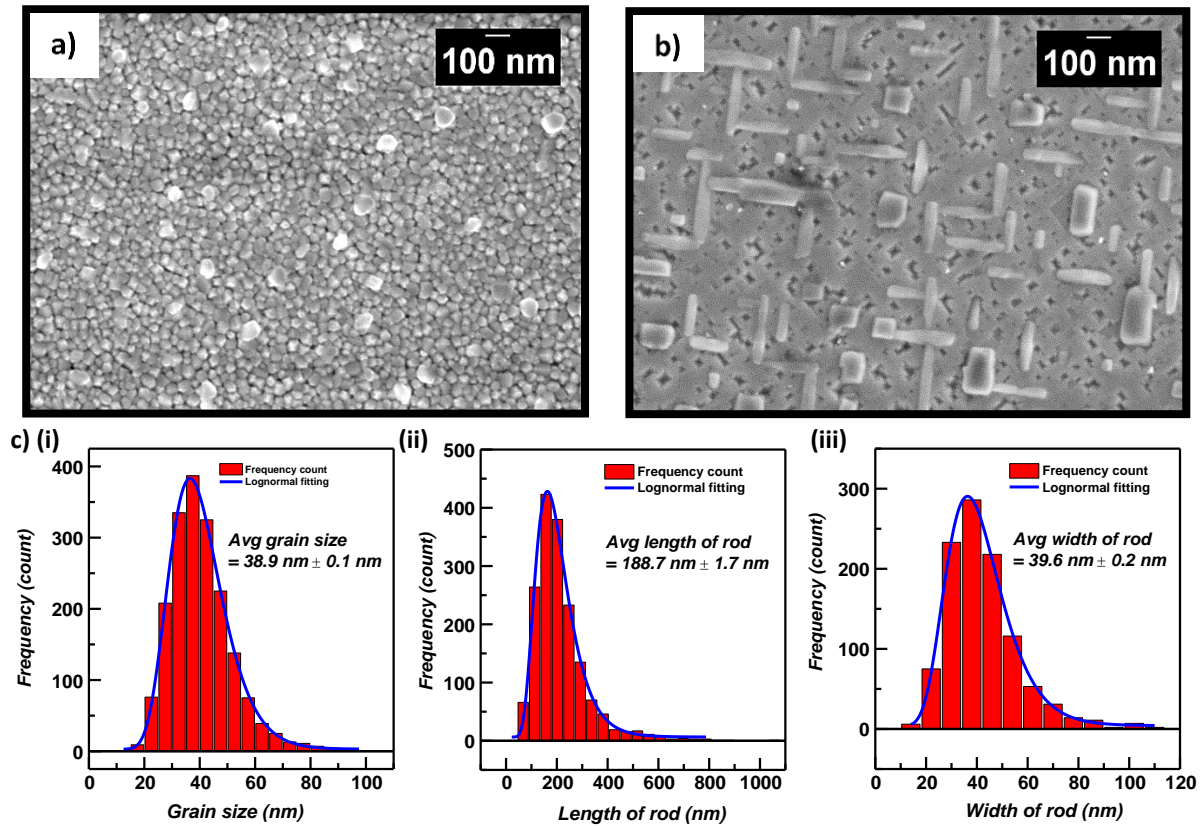

*Figure 1: Scanning electron microscopy images of NSMO thin films on STO. a) NS-G - granular morphology. b) NS-R – self-aligned-crossed-Nano-rod-morphology. c) The histograms illustrate the grain size calculation for NS-G and NS-R thin film. (i) Average grain size estimated for NS-G. (ii) Average rod length estimated for NS-R. (iii) Average rod width estimated for NS-R.*

Figure 1(a) shows the SEM image of NS-G thin films with granular morphology. The film is uniformly covered with multifaceted grains. Figure 1(c) (i) shows the average grain size estimated to be 38.9 nm. Figure 1(b) shows the SEM image of NS-R thin films with unique surface morphology. The thin film surface is uniformly covered with nano-rods crossed at right angles embedded in a matrix of NSMO containing square/rectangular pits. In NS-R, the average rod length is estimated to be 188 nm with an average width of 39.6 nm, as shown in the Figure 1(c), (ii) and (iii). Further, AFM measurements have been carried out on the NS-G and NS-R thin films. The 2D and 3D AFM scan in Figure 2 show columnar/island-type features in the NS-G thin film and crossed-rod features in the NS-R thin film.

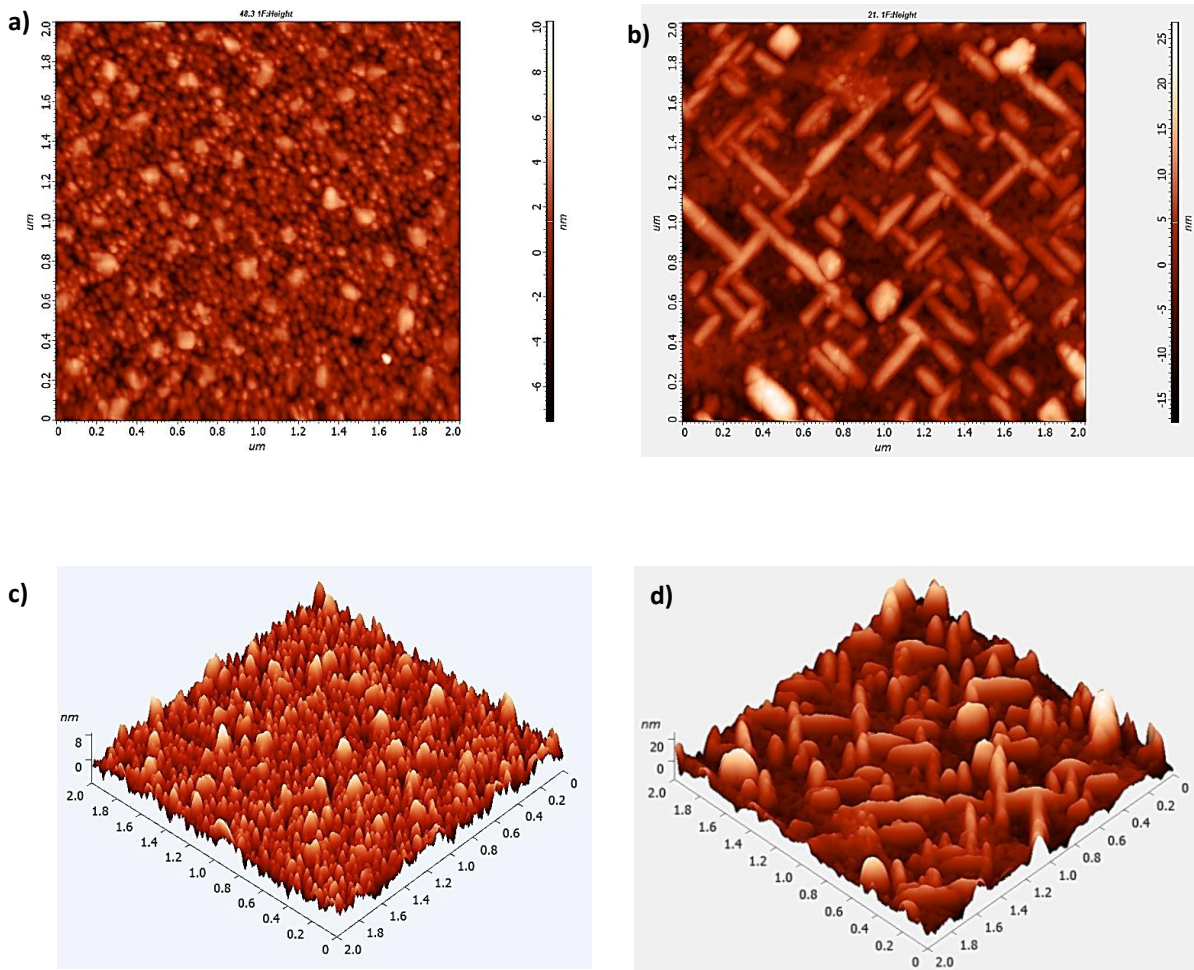

*Figure 2:* a*), b), 2D, and c), d) 3D AFM scans of grain-type NSMO thin film (left) and rod-type sample NSMO thin film on STO substrate.*

## 2. *Structural analysis of the thin film:*

The bulk NSMO compound has an orthorhombic crystal structure belonging to the *Pbnm* space group. In the pseudo-cubic (pc) representation, the unit cell parameter is given by $a_{pc} \approx c/2 \approx 3.849$ Å. The substrate STO has a cubic crystal structure with a lattice constant $a_{STO} = 3.905$ Å. NSMO grown on the STO substrate experiences a tensile strain due to the lattice mismatch of 1.4 %. The GI-XRD and high-resolution XRD (HR-XRD) reflections of the films are shown in Figure 3(a) and (b). The presence of multiple reflections in the GI-XRD scan of NS-G in Figure 3(a) reveals that the granular thin film is polycrystalline. In NS-R, the reflections of NSMO are absent, as seen in Figure 3(b). This may be due to its out-of-plane orientation with respect to the substrate. At the high 2θ angle ≈ 39.1°, the *(3 1 0)* STO plane gets aligned, resulting in high STO *(3 1 0)* reflection along with the NSMO *(2 4 0)* peak. This shows that the films are well-oriented, mirroring the substrate. Though NS-G is oriented, the crystallographic difference between NS-G and NS-R is attributed to the type of nano-structuring in the films.

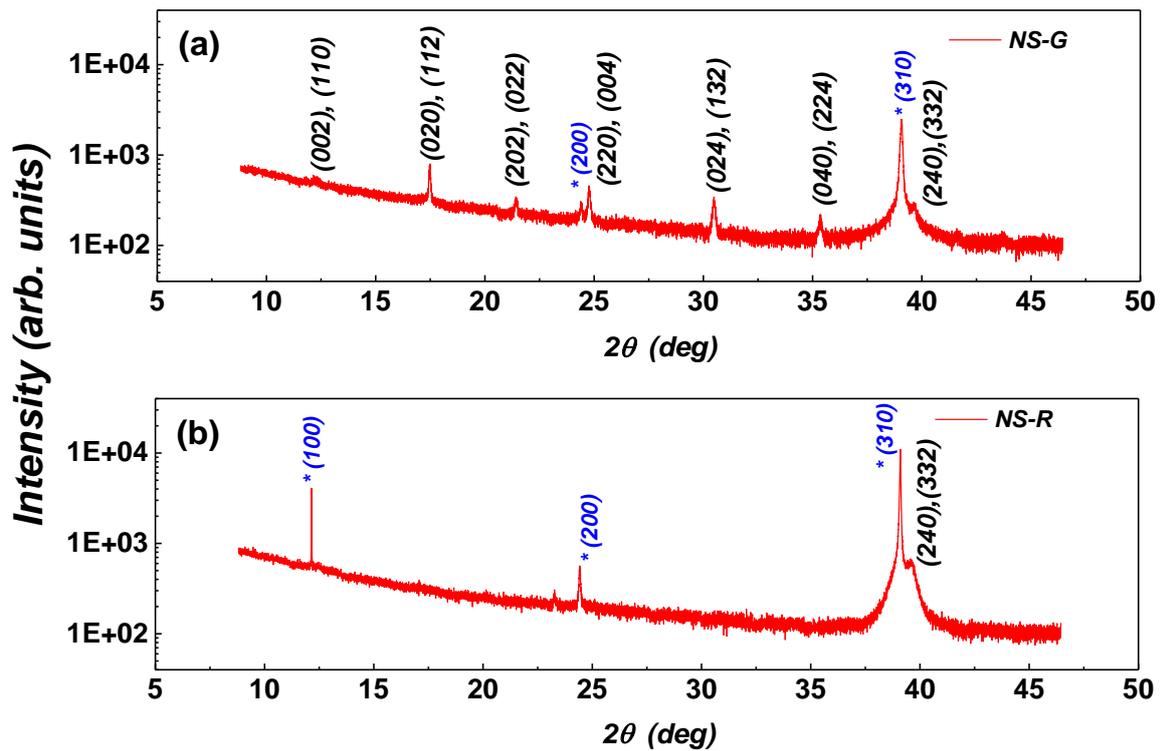

***Figure 3:*** *GI-XRD scans of NSMO thin films a) NS-G b) NS-R indexed using ICDD data (\* - STO peaks)*

### 3. *Effect of ex-situ annealing on morphology:*

To gain insight into the type of growth across these films, we compare the morphological changes in the in-situ annealed and ex-situ annealed samples in Figure 4. In granular thin films, no significant changes have been observed after in-situ and ex-situ annealing, apart from a minor increase in grain size, as seen in Figure 4(a). Whereas the sample with rod-type morphology obtained after ex-situ annealing in Figure 4(d) exhibits facetted droplets embedded in a matrix with rectangular holes and rod features in the in-situ annealed case, Figure 4(c). It is evident that once the initial growth mode is set, the ex-situ annealing aids in increasing grain size, relieving the strain in thin films in addition to decreasing oxygen defects in NSMO thin films[23]. We inspect the HR-XRD scans of the NS-G and NS-R thin films in the in-situ and ex-situ annealed cases to verify this claim.

**Pristine thin film after in-situ annealing**   **Further upon ex-situ annealing**

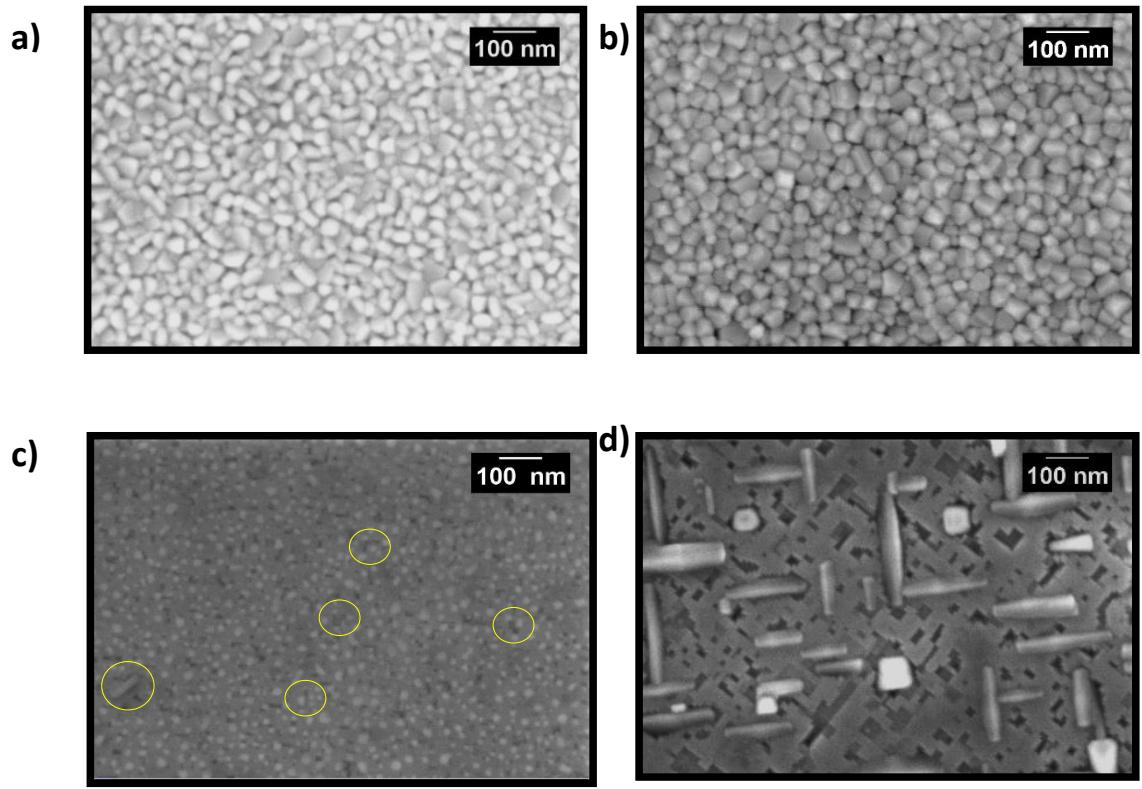

*Figure 4: Illustration of the effect of ex-situ annealing on NSMO thin films. a) SEM image of in-situ annealed granular thin film b) SEM image of the granular thin film after ex-situ annealing c) SEM image of the thin film after in-situ annealing showing rods and squared blocks in the encircled regions. d) SEM image of the same thin film after ex-situ annealing showing crossed-rod type morphology.*

Figure 5(a) and (b) show the HR-XRD scan performed over a range of 2θ (10º – 40º) for the films NS-G and NS-R after in-situ and ex-situ annealing. It is observed that the *(0 0 4)* NSMO peak is absent in the in-situ annealed NS-G thin film, whereas upon ex-situ annealing, NS-G shows improved texturing with the *(0 0 4)* NSMO peak close to the *(0 0 2)* substrate peak. In the case of NS-R, along with the substrate's (002) reflection, corresponding *(0 0 l)* pseudo-cubic reflections from NSMO are present with significant intensity even in the in-situ annealed condition. Further, as we compare HR-XRD scans of NS-G and NS-R after ex-situ annealing, the NS-R thin film has increased relative intensity compared with the NS-G.

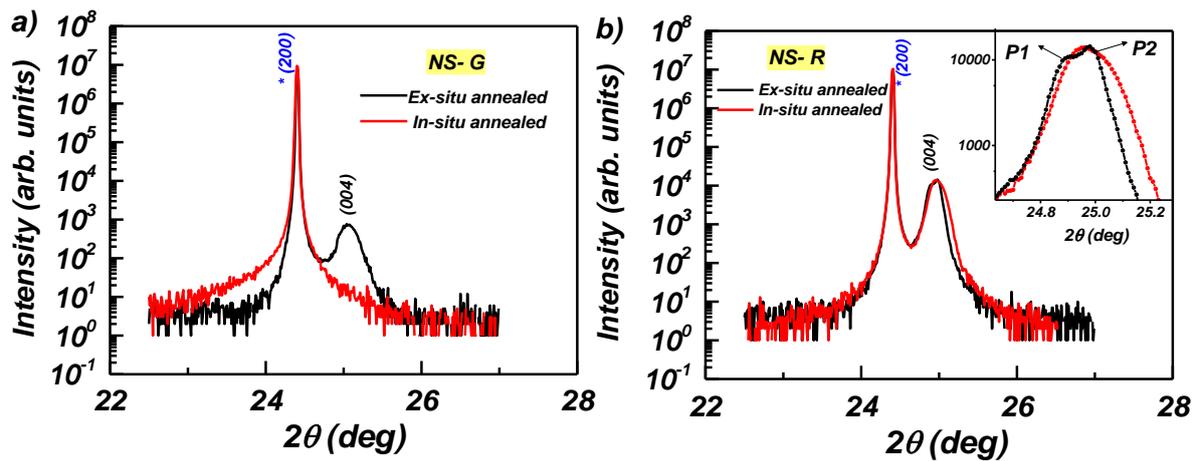

*Figure 5:* High resolution-XRD scan of NSMO thin films around the STO-(200) reflection inset: fine scan of NSMO (004) of NS-R sample showing double peaks – P1 and P2 (* - STO peaks)

Therefore, NS-R is highly oriented and more crystalline, which can be attributed to its epitaxial nature of growth. Additionally, the HRXRD scan of NS-R thin films after ex-situ annealing shows a doublet feature at its (004) reflection. A high-resolution fine scan was performed on the NS-R thin film to confirm the double peaks. Referring to the literature, we found that a similar doublet feature has been reported due to strain relaxation in PSMO thin films on STO substrate[31]. By fitting the peaks using the pseudo-Voigt function, as shown in figure S1 of supplementary information, the peaks were de-convoluted to evaluate the out-of-plane lattice parameter (tabulated in table T1 – supplementary information). The first peak was at $2\theta=24.88°$ with a c-lattice constant of 7.66 Å, and the second peak was at $2\theta=24.97°$ with a c-lattice constant of 7.63 Å. The reduction in the c-lattice constant of the second peak shows that there is compression of the lattice along the c-axis because of the tensile strain experienced by the thin film due to the substrate. Such a splitting in the peak was absent in films of thickness < 80 nm, indicating that this double peak is due to partial strain relaxation in the thicker film initiated by ex-situ annealing.

Thus, from the detailed XRD studies and discussions in the previous section, it is inferred that difference in initial-growth mode, and subsequent ex-situ annealing has prominently tuned the resulting surface morphology of the NSMO thin films. The granular thin film NS-G has multiple orientations similar to a polycrystalline system, whereas NS-R shows improved crystallinity and orientation mirroring the substrate. The parameters affecting the initial growth are discussed in the upcoming section.

## 4. *Effect of PLD parameters in tuning the morphology:*

PLD Parameters like laser energy density, oxygen partial pressure, and substrate temperature highly influence the type of growth. Changes in these parameters lead to variations in the energy of the ad-atoms deposited on the substrate. To understand the role of $O_2$ partial pressure and laser energy density during the deposition, we have prepared NSMO thin films by varying these parameters. Post deposition, the films were in-situ annealed at 750 °C for 2h in an oxygen background pressure of 1 bar. Ex-situ annealing was carried out subsequently.

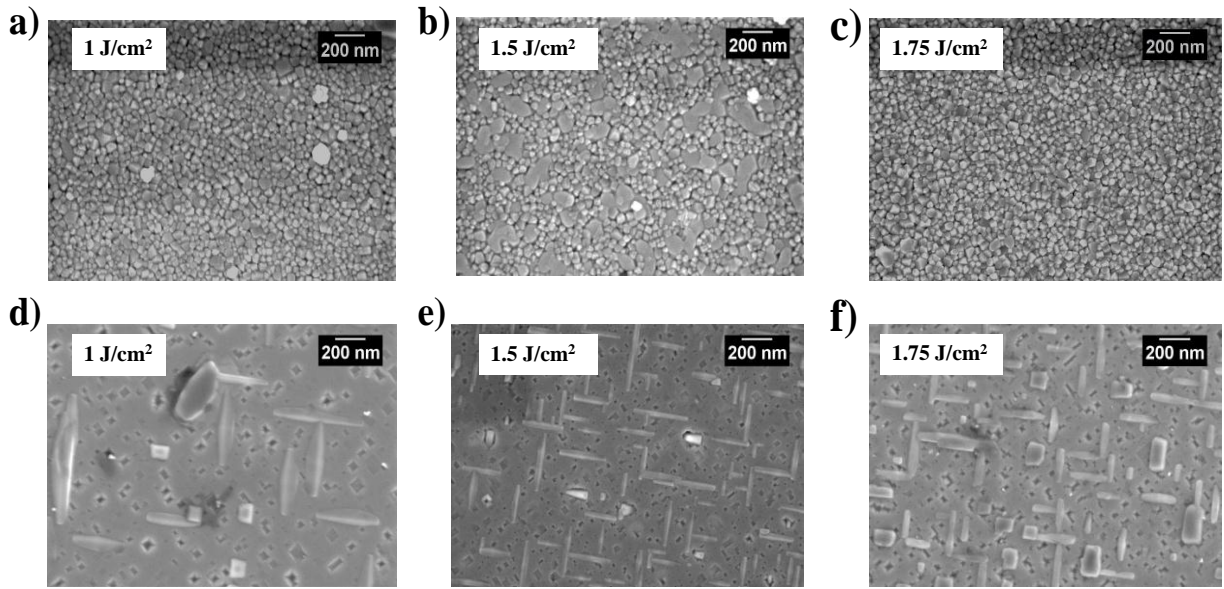

***Figure 6:*** *The SEM images of NSMO thin films with granular morphology (a), (b), and (c) and rod morphology from (d), (e), and (f) obtained at corresponding laser energy density- 1 J/cm$^2$, 1.5 J/cm$^2$, and 1.75 J/cm$^2$.*

Figure 6 presents the morphology of films deposited under different laser energy densities varied from 1 to 1.75 J/cm$^2$. During deposition, the oxygen partial pressure and substrate temperature were maintained at 0.36 mbar and 750 °C. Figure 7 presents the morphology of films obtained at different oxygen partial pressure of 0.3 mbar, 0.4 mbar, and 0.5 mbar while the laser energy density and substrate temperature were maintained at 1 J/cm$^2$ and 750 °C during deposition, respectively.

We found that changes in oxygen partial pressure and laser energy density did not influence the surface morphology, as both type of morphologies have been observed in different deposition runs with the same parameters. Further, as we have obtained granular and rod-type films for the same substrate temperature of 750 °C, the role of substrate temperature is also ruled out. Thus, irrespective of changes in the parameters mentioned above, thin films of either granular or crossed-rod nanostructure were obtained. Therefore we suspect the substrate and the strain it offers to plays a vital role in altering the growth mode of the thin film.

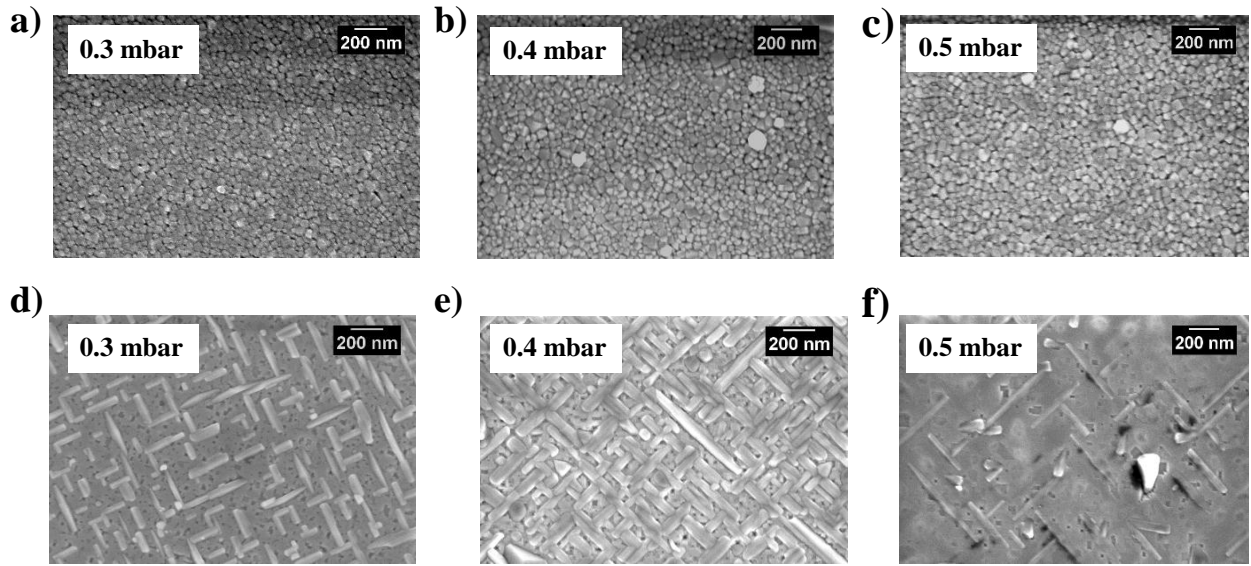

*Figure 7: The SEM images of NSMO thin films prepared at oxygen partial pressure of 0.3 mbar, 0.4 mbar, and 0.5 mbar. a), b), and c) are granular NSMO thin, and films with rod morphology are shown in d), e), and f) at corresponding oxygen partial pressure.*

### 5. *Effect of miscut angle in tuning the morphology:*

The commercial STO substrates used here are one-sided polished, and their surface was found to have a miscut. In commercially purchased wafers, the occurrence of a miscut in the range of 0.05°-0.3° is well known and unavoidable due to mechanical cutting and polishing of single crystal STO wafers[14,32]. In Figure 7(a), the as-received STO substrate, after cleaning, shows clear terrace features in the AFM scan, confirming the presence of miscut on the substrate surface. In a given wafer, the miscut can be in-plane or out-of-plane or both (some works refer to this as miscut directions φ and θ instead of in-plane and out-of-plane, respectively). The miscut angle and direction can alter the growth mode as the lattice strain is anisotropic along the substrate surface and step edges[6], thus resulting in different surface morphology by forming anisotropic structural domains[33]. Several works are available in literature [33–35] on the growth of manganite thin film on STO substrate with miscut. These reports claim that the value of the miscut angle and appropriate adjustments in growth conditions can control the number of structural domains in the thin film. As we have already ruled out the possibility of growth conditions influencing the resulting morphology, we tried to evaluate the value of miscut present in our STO substrates to see if it has affected the resulting morphology.

To determine the value of miscut present in the substrates, we have followed the XRD-protocol from literature[36]. This was carried out in a BRUKER D8, Lab source XRD setup. According to the protocol, a low incident angle (~0.2°) rocking-scan was initially performed to ensure that the sample was aligned with the X-ray. This was done to optimize the angle of the sample holder, and the offset in the 2θ value (~0.4°) was noted as ζ. Following that, a rocking scan was performed around the (200) peak of STO (46.483°), and phi & chi scans were done to orient the wafer. Further, the rocking scan around the (200) peak of STO was repeated, fixing the X-ray tube position. Finally, a detector scan was performed around the (200) peak of STO and this time the offset in 2θ was noted as ζ'. The difference δζ, between ζ and ζ', gives the estimate of miscut. Next, the sample was rotated by 90°, and the scans mentioned above are repeated in the same order. The difference between the offsets obtained this time was denoted as δξ. Finally, the out-of-plane miscut angle was evaluated using equation (1). After determining miscut on various STO wafers, we found that the value out of plane miscut angle varies from 0.13° up to 0.48°.

$$\theta_{out-of-plane} = \arctan\sqrt{\tan^2(\delta\zeta) + \tan^2(\delta\xi)} \tag{1}$$

| Sample | Miscut angle | Granular Morphology | Sample | Miscut angle | Rod type Morphology |
|---|---|---|---|---|---|
| NS-G | 0.48° | 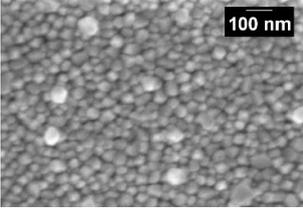 | NS-R | 0.31° | 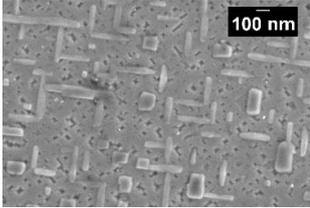 |
| G1 | 0.31° | 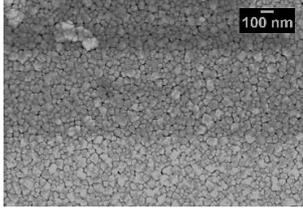 | R1 | 0.19° | 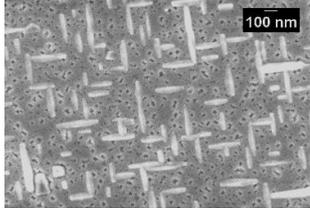 |
| G2 | 0.30° | 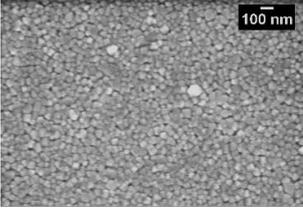 | R2 | 0.25° | 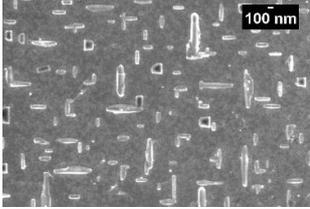 |

*Table 1 : This table illustrates the morphology of NSMO thin films obtained on STO substrates with different values of miscut.*

We see from table T1 that, both granular and rod-type morphology was observed on substrates with miscut angle varying from 0.13º up to 0.48º. Sample G1 with granular morphology and NS-R with rod-type morphology, possess the same miscut angle of ~ 0.3º. This is very interesting, as the value of the miscut angle has not influenced the altered growth modes present in our samples. Therefore to comprehend the resulting morphology, we have further investigated the type of growth occurring on the terraced surface.

### 6. *Thin film growth on the terraced surface:*

A miscut on the substrate is useful for epitaxial thin films[37] as the steps and terrace edges act as nucleation centres and result in a step flow growth mode[38]. But the actual processes governing the step-flow growth are more complex. The basic parameters driving this type of growth are the coefficient of diffusion and the height of the Ehrlich-Schwoebel (ES) barrier[39]. The diffusion of the adatoms on the surface and their incorporation into the crystal structure govern the formation of different morphologies at the surface. Additionally, the ES barrier at the terrace/step edges introduces an asymmetry in the potential energy at the edge. An adatom, reaching the terrace, either nucleates or descends into the step depending on the ES barrier height. Similarly, an adatom reaching below the step experiences an inverse step barrier which prevents the particles from attaching to the step from below.

If the barrier height is appropriate, ad-atoms can properly attach themselves to the step edges resulting in a step flow growth. However, the existence of the barrier makes the growth on the stepped surfaces highly unstable resulting in modified surface features such as step meandering, nano-columns/wire formation, spirals/mound formations, and faceted pits. In a recent work by Magdalena et al.[40], a simulation using the Cellular Automaton model in (2+1)D gave rise to different patterns of surface morphology on vicinal surfaces. According to the simulation, different processes occurred depending on the values assigned to the barrier height at step edges. The adatom could either attach to the step to build the crystal by jumping/ descending at the step edge or scatter away from the barrier resulting in the formation of islands. For a fixed adatom flux, diffusion of adatoms takes place on the vicinal surface, and probabilities are assigned for each of the processes mentioned above. Depending on the probability value, various surface patterns were simulated for three cases. In case (i), for a high ES barrier, the three-dimensional surface formation resulted in square/rectangular islands following the cubic lattice symmetry at the middle of the terraces. In case (ii), with a reduced

ES barrier height, more atoms were trapped at the top of the step, and a new pattern of nanocolumns emerged consisting of cubic formations with deep narrow cubic pits. Finally, in case (iii), when the height of the barrier was adjusted such that the probability of the adatoms descending the step is equal/of the same order as the probability of the adatoms jumping up to the step from below, it resulted with nano-wire or a columnar growth. Further, the presence of additional local sinks that alters the potential barrier also resulted in nano-columns/islands at random positions.

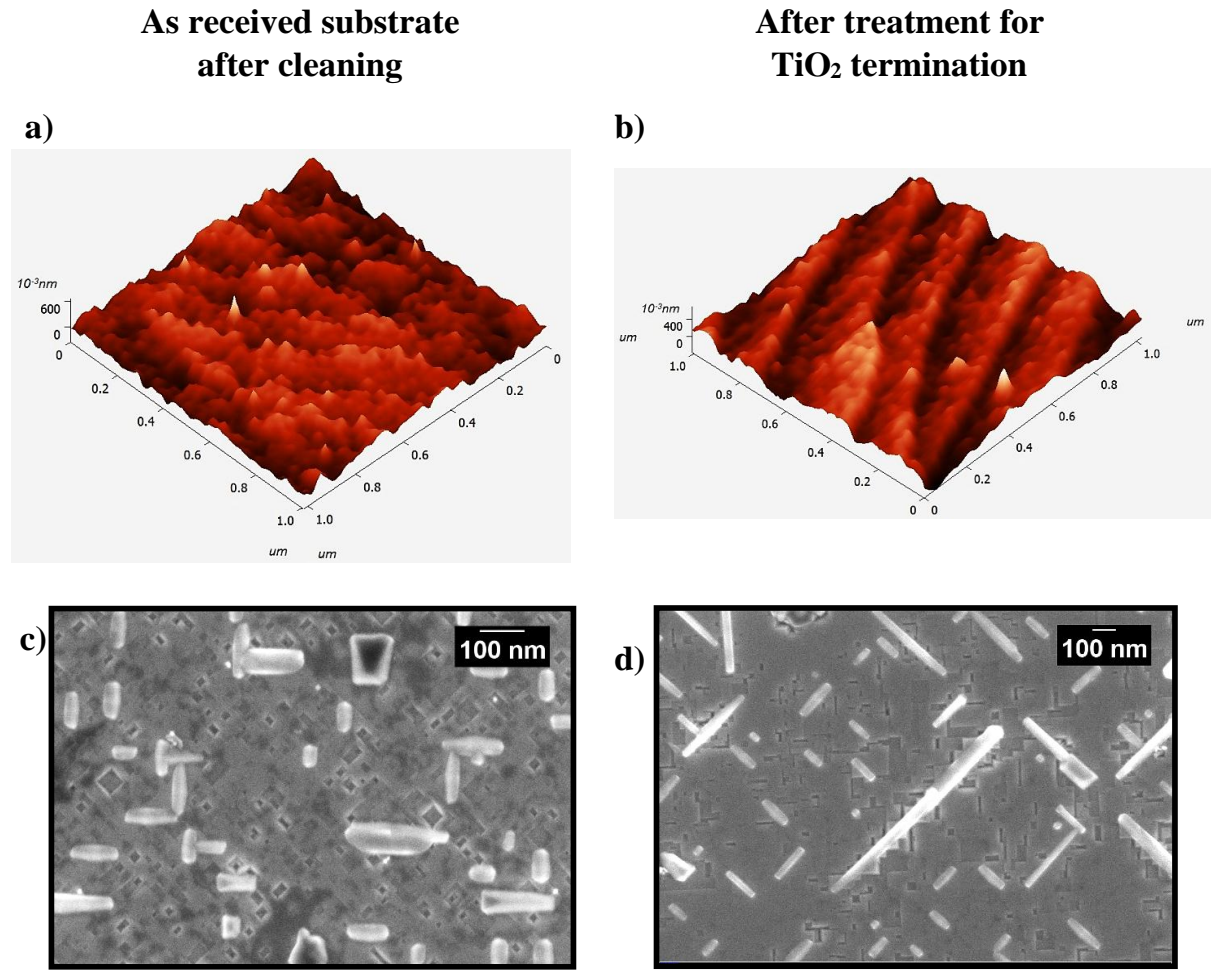

*Figure 8:  AFM scan of the STO substrate a) as-received commercial substrate after cleaning b) the same substrate after TiO$_2$ termination obtained after heat treatment method with a step height of ~ 0.4 Å (one-unit cell height of STO). c), d) NSMO thin films grown on the corresponding substrates*

Thus, we can understand that our resulting granular morphology on the miscut STO substrate is precisely similar to the surface morphology resulting from the case (iii). In the STO susbtrate, the presence of disoriented terraces and improperly removed SRO terminations may have altered the ES barrier resulting in local sinks at the substrate surface, thus resulting in the island/columnar growth. Finally, the surface morphology of the NS-R thin film resembles the morphology they obtained in case (ii). This fact can be verified from a close inspection of the surface of NS-R at high magnification in Figure 9(a). The surface morphology clearly shows layer-by-layer growth with squared pits.

Further, in attempting to reduce the local sinks, NSMO thin films were synthesized on pure $TiO_2$ terminated substrates. The substrates are treated with DI water and then annealed at high temperatures according to the protocol for $TiO_2$-termination[41]. The treatment produced clear step and terrace characteristics in the substrate, as observed in the AFM scan shown in Figure 8(b). NSMO thin films were deposited on these substrates and, subsequently, ex-situ annealed. The SEM imaging revealed that they exhibited similar rod-type morphology where the rods are self-aligned and crossed at right angles embedded in a matrix of NSMO with rectangular features, shown in figure 8(d). This procedure was repeated on several $TiO_2$-terminated STO substrates, and we could reproduce the same morphology. This is because the complete removal of SrO assures the absence of local sinks and suppresses the island/columnar growth. However, rods in the thin film are believed to arise from droplets deposited due to high laser energy density (1.75 $J/cm^2$). This is verified in the SEM images of in-situ annealed NSMO thin film shown in Figure 4(b), where the droplets are elongated into rods upon ex-situ annealing.

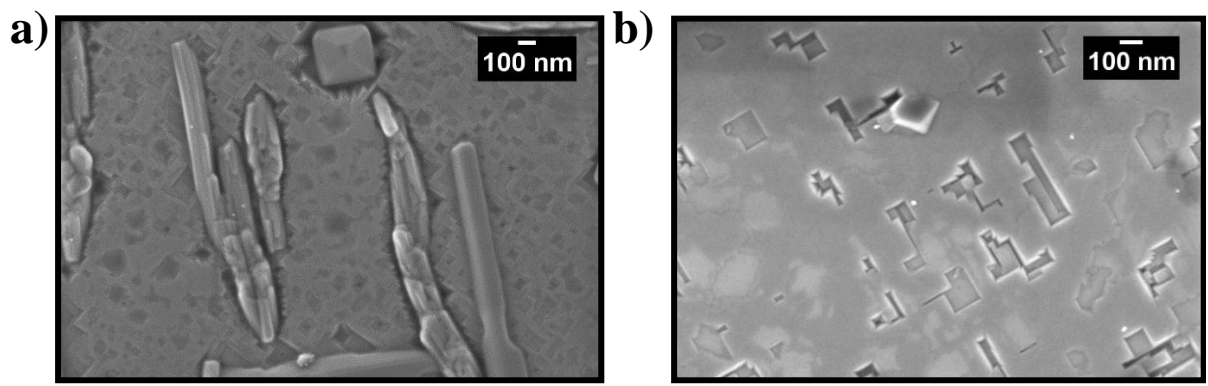

*Figure 9: SEM images of NSMO thin films under high magnification. a) NSMO thin film grown on as-received, cleaned STO substrate deposited b) NSMO thin film grown with laser fluence on $TiO_2$ terminated STO substrate*

Lastly, to obtain smoother films, we have synthesized NSMO thin films on fully $TiO_2$ terminated STO substrate at low laser energy density (1 $J/cm^2$), reducing droplets' density. As expected, we obtained thin films with reduced density of rods with the same type of morphology. The SEM image of the film is shown in Figure 9(b), free of nano-rods. The films have rectangular faceted pits, and layer-by-layer growth is evident through the holes.

Therefore the ES barrier plays a significant role in vicinal surfaces and can result in the spontaneous ordering of adatoms resulting in unique surface nanostructures. Thus we emphasize that when films are grown on a commercial substrate, the resulting morphology can be either granular or rod-type depending on the potential energy landscape that depends upon a wide range of parameters, including the size, shape of terraces, and type of terminations present at the substrate.

7. *Electrical-transport measurements:*

The nanostructure plays a vital role in the transport behaviour of a manganite thin film system[30]. To understand the transport behaviour of the nanostructured NSMO thin films, the resistivity measurements are carried out using the standard 4-probe geometry [42] and plotted as a function of temperature in Figure 10. It is observed that the granular film NS-G has higher resistivity as compared to NS-R. Both thin films, NS-G and NS-R, exhibit the insulator-to-metal transition (MIT), and the transition temperature $T_{MIT}$ is found to be 147 K for sample N-G and 135 K for NS-R. The transition into the metallic regime is sharper in the case of NS-R compared to NS-G thin film. The electrical transport behaviour has been analysed using different theoretical models and fitted in the corresponding temperature regimes. The best fit in each region is chosen based on the reduced $\chi^2$ value.

$$\rho(T) = \rho_R T \exp\left(E_a / K_B T\right) \qquad \ldots(2)$$

$$\rho(T) = \rho_0 \exp\left(T_0 / T\right)^{1/4} \qquad \ldots(3)$$

$$E_{hopping} = \frac{\kappa_B T_0^{1/4} T^{3/4}}{4} \qquad \ldots(4)$$

The high-temperature insulating phase is studied using the small polaron hopping (SPH) model and the variable range hopping (VRH) mechanism given by equations (2) and (3), and hopping energy is calculated from equation (4) [42,43]. The VRH model better fits the high-temperature region ($\approx$ 195 K to 300 K) for both films. The hopping energy in the case of NS-G is 128 meV and 125 meV for NS-R, in agreement with the order of value reported for manganite thin films (~100 meV) [43,44]. The resistivity in the metallic region below $T_{MIT}$ is generally fitted with an empirical equation (5). At low temperatures, in addition to the temperature-independent scattering effects from defects and grain boundaries (GBs) ($\rho_o$), etc., scattering effects due to electron-electron ($\rho_2$), electron-magnon ($\rho_{4.5}$) and electron-phonon ($\rho_P$) dominate along with the strong correlation effects ($\rho_{0.5}$) [45].

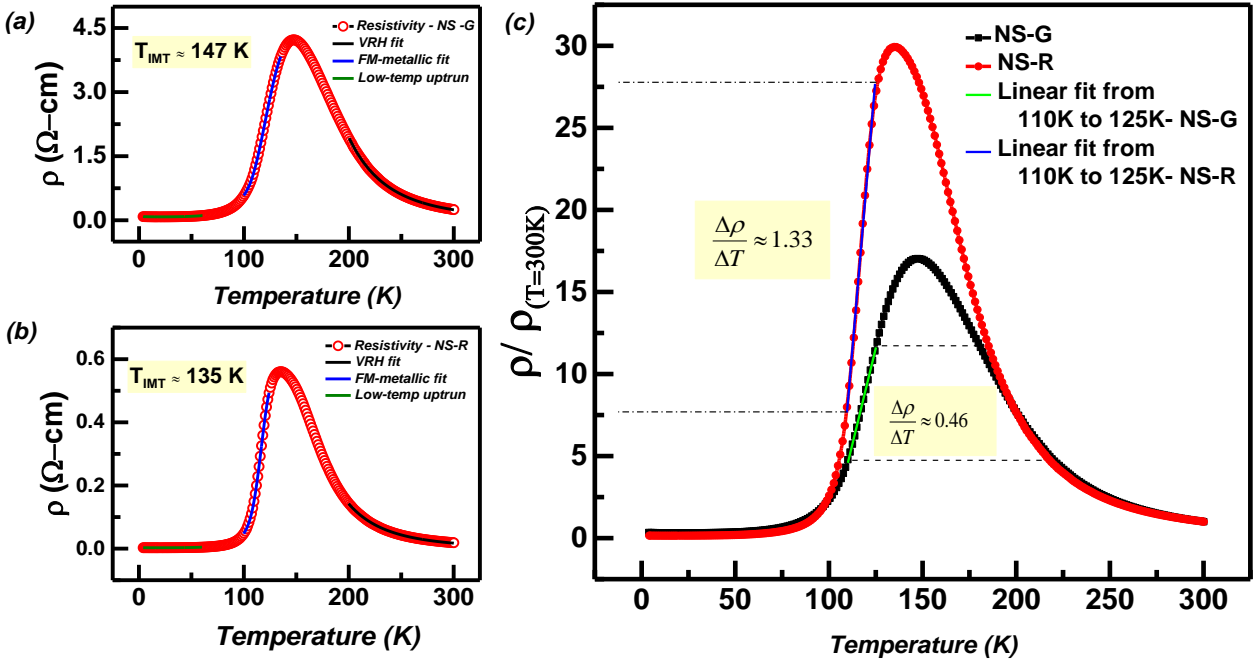

***Figure 10:*** *a), b) Resistivity vs. temperature curve of NSMO thin films – NS-G and NS-R showing insulator to metal transition with decreasing temperature and fitted according to theoretical models in different temperature regimes c) Normalized resistivity plot of NS-G and NS-R thin film. Inset: Plot of variation of TCR with respect to temperature.*

A low-temperature resistive upturn is observed below 50 K in both films. In figure S3 of supplementary information, the resistive upturn in the low-temperature region from 4 K up to 60 K is fitted using equation (6), which considers all the scattering mechanisms mentioned above. An enhanced resistive upturn is observed at low temperatures in NS-G. This is due to the enhanced GB-scattering effect and the contribution from other scattering mechanisms at low-temperature. The contributions from different scattering mechanisms are analysed, and the values are tabulated in supplementary information Table T2.

The intermediate temperature regime from 90 K to 134 K in the ferromagnetic-metallic state is fitted using the equation (7). The addition of the polaronic term to the resistivity gives a better fitting in this region as theoretical models claim the formation of polaron near the MIT[46].

$$\rho(T) = \rho_o + \rho_m T^m \tag{5}$$

$$\rho(T) = \rho_o + \rho_2 T^2 + \rho_{4.5} T^{4.5} + \rho_P T^5 + \rho_{0.5} T^{0.5} \tag{6}$$

$$\rho(T) = \rho_o + \rho_2 T^2 + \rho_{4.5} T^{4.5} + \rho_P T^5 + \rho_{0.5} T^{0.5} + \rho_{7.5} T^{7.5} \tag{7}$$

$$TCR\% = \frac{1}{\rho}\left(\frac{d\rho}{dT}\right) \times 100 \tag{8}$$

An interesting feature is observed in the resistivity plots of the NS-G and NS-R thin films apart from the low-temperature resistive upturn. In Figure 10(c), the resistivity of both the thin films (NS-G and NS-R) has been normalized with their resistivity at 300 K, and a linear fitting in the metallic region below $T_{MIT}$ (110 K to 125 K) is carried out to determine the slope. The resistivity slope of samples with rod morphology differs from samples with granular morphology up to an order of magnitude. The increase in slope value below the transition temperature indicates the sharpness of the resistive transition for the samples with rod morphology. This characteristic increase in slope up to an order is evident in all our thin films with rod-type morphology (see supplementary figure S2). To characterize the sensitivity of resistance with respect to changes in temperature, the temperature-coefficient of resistance (TCR) has been evaluated using equation (7). It was found that NS-R has a higher value of TCR %, ~ 12 %, compared to NS-G with TCR %, ~ 7 %. Additionally, the samples with rod morphology were found to have enhanced TCR% (supplementary information figure S2). To comprehend this result, we discuss the effect of GBs on the conduction mechanism.

The manganite system undergoes a disorder-induced phase transition from PM to FM state with decreasing temperature[21]. Due to phase co-existence during the transition, the conduction channel is presumed to have filamentary FM paths in the PM matrix [47]. Conduction takes place through the percolation of current across the well-connected FM regions. In addition to the FM filamentary path, the GBs also play a significant role in the conduction mechanism. We refer to Verutruyen *et al.*'s [48] work which explores the effect of a single GB in the La-Ca-Mn-O (LCMO) system. They showed that the resistivity falls sharply at the transition temperature when measured on a single grain of LCMO (free of GBs). However, when measured across a single GB, the resistivity initially decreased, followed by a broad resistive feature near the transition temperature. Thus, in a granular system, though the conduction takes place through the percolation paths of well-connected FM regions, the GBs cause increased resistivity due to increased spin-dependent scattering across the GB[47]. The above explanation is consistent with our results, where the thin film with granular morphology (NS-G) shows a broad resistive transition below the transition temperature with reduced TCR

%. If the connectivity is enhanced between the grains, a sharper decrease in the resistivity can occur in the metallic regime. Remarkably, we observe that all of our thin films with rod-morphology show sharp resistive transition near MIT irrespective of the thickness of the film. Thus, this nanostructure aids improved conduction in the FM metallic phase, leading to the sharp resistive transition with enhanced TCR % comparable to that of a highly-crystalline system. Attempts to enhance the TCR % have been carried out by doping with elements such as Ag, as high TCR % is required for applications in sensors and infrared detectors[49,50]. These elements precipitate as nanocomposite in the manganite system and improve the conductivity, leading to a sharper resistive transition. However, in our study, we have substantiated that the enhancement of TCR % is possible with proper tuning of the nanostructured morphology of thin films.

**CONCLUSION:**

In conclusion, the PLD-grown NSMO thin films were observed to have two prominent surface morphologies – granular and crossed-nano rods. The metal-to-insulator transition (MIT) temperature, $T_{MIT}$, was found to be 147 K for a granular NSMO (NS-G) thin film and 135 K for a thin film with crossed-rod morphology (NS-R). The nature of the resistive transition is broad in the former, whereas the latter exhibits a sharp MIT feature. The temperature coefficient of resistance (TCR) was evaluated, and NS-R thin film has a higher value of TCR %, ~ 12 %, compared to NS-G with TCR % ~ 7 %. Additionally, we have observed that all the films with rod-type morphology exhibit a significant enhancement in TCR% up to one order of magnitude compared to the granular thin film. Thus, we have demonstrated that TCR % can be enhanced with proper tuning of the nanostructures in thin films, which is relevant for technological applications. The reason for such nano-structuring is explored in great detail. It was found that parameters like laser energy density, $O_2$ partial pressure, and the substrate miscut angle had minimal effect. At the same time, the difference in the potential landscape of the Ehrlich-Schwoebel (ES) barrier is believed to play a vital role in the growth dynamics of the films. Films grown with reduced laser energy density (1 J/cm$^2$) on the TiO$_2$ terminated substrates exhibited highly reproducible layer-by-layer growth. This substantiates the presence of reduced local sinks and ES barrier height, resulting in epitaxial growth of NSMO thin films. Therefore, a fine-tuning of a wide range of parameters, including strain and surface terminations, is required to obtain a fine control of the ES barrier that influences the growth process of thin films. This paves the way for investigation into the role of the ES barrier in manganite thin film growth. Using RHEED and in-situ STM techniques, a few groups have already attempted to experimentally determine the value of the ES barrier on SrTiO$_3$ substrates for the growth of La-Ca-Mn-O manganite system[51]. It would be interesting to explore the relationship between the value of the ES-barrier and the type of morphology experimentally in the future.

**Author contributions**

The division of work is as follows: NSMO thin film samples were prepared by R.S.M. SEM imaging was carried out by S.A, J.P. AFM measurements were carried out by K.G. XRD measurements were carried out by R.M.S, P.N.R, PG, S.K.R. Magneto-transport measurements were carried out by R.S.M and E.P.A. Analysis were done by R.S.M, E.P.A, and S.A. Writing was carried out by R.S.M, and all authors discussed the results and commented on the manuscript. E.P.A., T.G.K and A.M. supervised this research work.

**Conflict of interest:**

The authors declare no conflict of interest.

**Acknowledgments**

One of the authors (R S Mrinaleni) would like to acknowledge the Department of Atomic Energy, India for the provision of experimental facilities. We thank UGC-DAE CSR, Kalpakkam node, for providing access to magnetic and magnetotransport measurement systems. The authors are grateful to RRCAT, Indore, for beam line facilities.

**Funding statement:**

One of the authors (R S Mrinaleni) would like to acknowledge the funding support from the Department of Atomic Energy, India.

49. Li, J. *et al.* Improvement of electrical and magnetic properties in La0.67Ca0.33Mn0.97Co0.03O3 ceramic by Ag doping. *Ceram. Int.* (2022) doi:10.1016/j.ceramint.2022.08.255.

50. Jin, F. *et al.* La0.7Ca0.3MnO3-δ:Ag nanocomposite thin films with large temperature coefficient of resistance (TCR). *J. Mater.* (2022) doi:10.1016/j.jmat.2022.01.010.

51. Gianfrancesco, A. G., Tselev, A., Baddorf, A. P., Kalinin, S. V & Vasudevan, R. K. The Ehrlich–Schwoebel barrier on an oxide surface: a combined Monte-Carlo and in situ scanning tunneling microscopy approach. *Nanotechnology* **26**, 455705 (2015).

# *SUPPLEMENTARY INFORMATION*

## I- Deconvolution of NSMO (004) reflection:

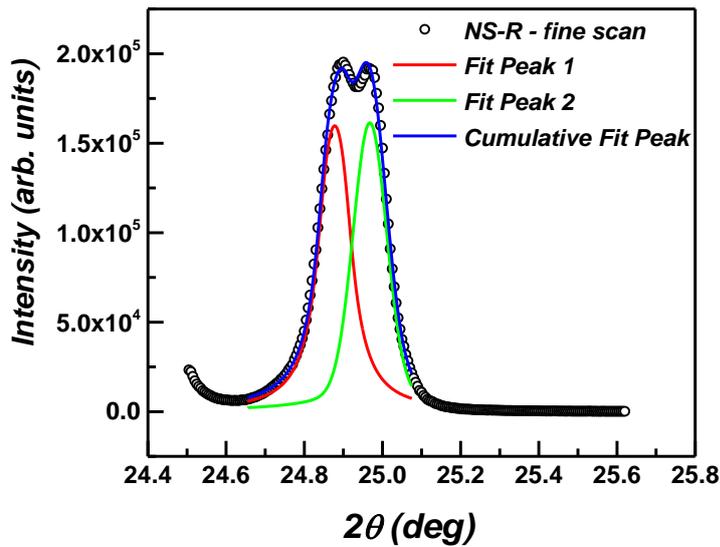

*Figure S1: The double peak in the HR-XRD scan of NS-R thin film is confirmed by a HR-fine scan. The individual peak positions are noted as the centre of the fitted peaks P1 and P2.*

| Sample | 2θ for (004) reflection (° deg) | Calculated c-lattice parameter (Å) |
|---|---|---|
| NS-G | 25.05° | 7.61 |
| NS-R | 24.88° – P1 | 7.66 |
|  | 24.97° – P2 | 7.63 |

Table ST1: The table illustrates the values of c-lattice parameters evaluated from the (004) NSMO reflection.

## II- Transport studies on NSMO thin films

Three samples with granular morphology G-A, G-B, G-C, and rod morphology R-A, R-B, R-C, were selected and their resistivity was measured using 4-probe technique. The normalized-resistivity plot for the selected NSMO thin films are shown Figure. 4. The value of resistivity is different across the NSMO thin films, since they are deposited under slightly different PLD conditions but all of them exhibited MIT. Observing the nature of MIT transition in these selected samples, G-A, G-B, G-C with granular morphology have a broad resistive transition below their MIT temperature. The samples R-A, R-B, R-C with rod-morphology show a sharp resistive transition in the FM-metallic state below their MIT temperature. The value of slope is evaluated from the linear fit in the metallic region and it shows that samples with rod-type morphology have increased slope up to one order as compared to the granular samples. Temperature coefficient of resistance (TCR) is evaluated for these films and it is found that samples G-A, G-B, G-C have peak-TCR % of 5 %, 4 %, and 8 % at 105 K, 77 K, and 121 K, respectively. An enhanced $TCR_{peak}$ % is obtained for samples with rod-morphology. The samples R-A, R-B, R-C have peak-TCR % of 21 %, 14.5 %, and 18 % at 98 K, 80 K, and 100 K, respectively.

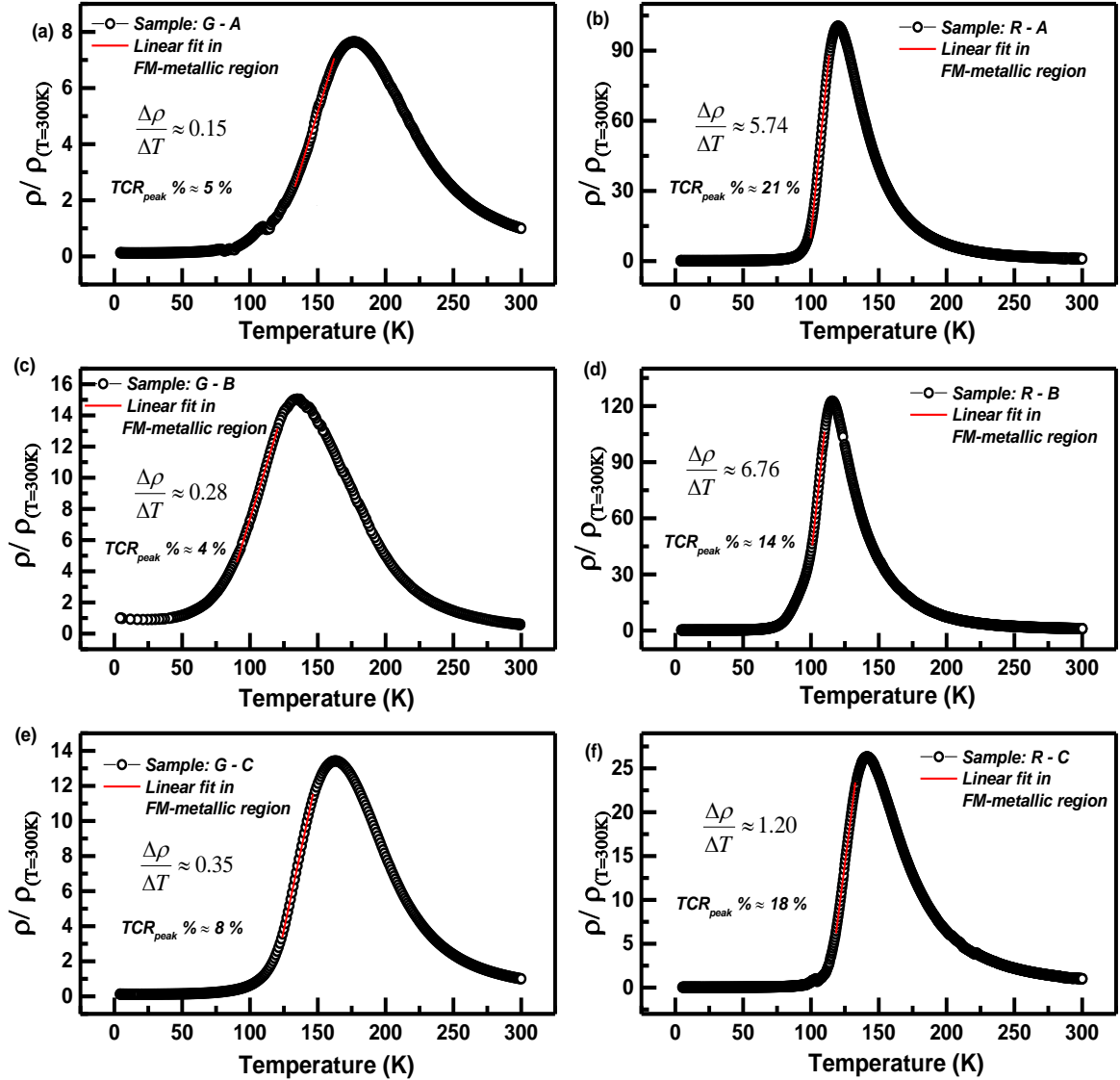

*Figure S2: Plots of normalized resistivity vs. temperature of NSMO films with granular and rod-type morphology. (a),(c),(e): Samples with granular morphology G-A, G-B, G-C. (b),(d),(f): Samples with rod morphology R-A, R-B, R-C. A linear fit in the FM-metallic region give the rate of change of resistivity with respect to temperature.*

## III- Low-temperature studies on NSMO thin films – NS-G and NS-R

To study the low-temperature transport across the thin films with different morphology, the plot of low-temperature resistivity of the granular thin film NS-G and rod-type thin film NS-R is shown in figure S5. An enhanced low-temperature resistive upturn is observed in NS-G from figure. S5. Using the low-temperature transport equation the resistivity data is fit and the fitting parameters are summarized in table ST2. The first term, $\rho_o$ which represents the contribution from grain-boundary (GB) scattering is found to be higher by more than one-order in NS-G as compared to NS-R. This is expected as NS-G has a granular morphology and increased contribution from GB scattering affects the transport mechanism even at low-temperatures. Additionally, $\rho_o$'s value is higher by orders of magnitude as compared to the other coefficients. This shows that GB scattering effects dominate the transport mechanism compared to other contributions to the electronic transport.

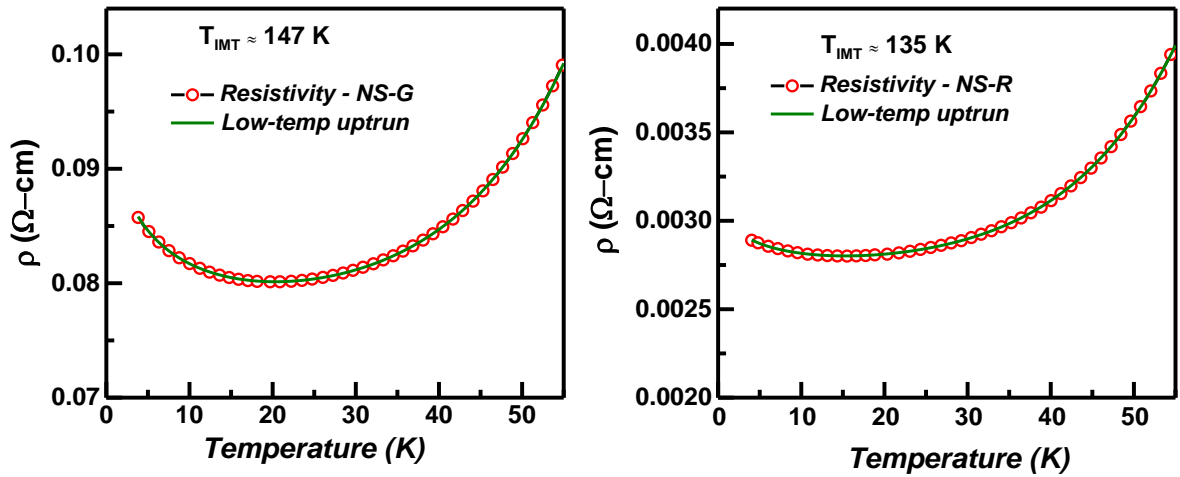

*Figure S3: Low-temperature resistive up-turn is observed in the NSMO thin films NS-G and NS-R. The temperature regime from 4 K up to 60K is fit using the low-temperature transport equation.*

| Sample | $\rho_o$ | $\rho_2$ | $\rho_{4.5}$ | $\rho_P$ | $\rho_{0.5}$ | $R^2$ (%) |
|---|---|---|---|---|---|---|
| NS-G | 0.09272 | 1.16E-5 | -1.00E-9 | 1.33E-10 | -0.0038 | 99.99 |
| NS-R | 0.00305 | 3.53E-7 | -2.90E-11 | 4.90E-12 | -8.38E-5 | 99.99 |

**Table ST2: The table illustrates the values of coefficients of low-temperature transport after fitting.**